\input{aipcheck.tex}
\documentclass[
    ,final            
  ]
  {aipproc}

\layoutstyle{6x9}

\begin{document}

\title{The gamma-ray emitting microquasar LS~I~+61~303}


\author{M. Massi}{
  address={Max Planck Institut f\"ur Radioastronomie, Auf dem H\"ugel 69,
D-53121 Bonn, Germany},
  email={mmassi@mpifr-bonn.mpg.de},
  thanks={}
}

\author{M. Rib\'o}{
  address={Service d'Astrophysique, CEA Saclay, B\^at. 709, L'Orme des Merisiers, 91191 Gif-sur-Yvette, Cedex, France},
  email={mribo@discovery.saclay.cea.fr},
}

\author{J.~M. Paredes}{
  address={Departament d'Astronomia i Meteorologia, Universitat de Barcelona, Av. Diagonal 647, 08028 Barcelona, Spain},
  email={jmparedes@ub.edu},
}
\author{S.~T.\ Garrington}{
  address={Nuffield Radio Astronomy Laboratories, Jodrell Bank, Macclesfield, Cheshire SK11 9DL, UK},
  email={stg@jb.man.ac.uk},
}
\author{M. Peracaula}{
  address={Institut d'Inform\`atica i Aplicacions, Universitat de Girona, Campus de Montilivi s/n, 17071 Girona, Spain},
  email={marta.peracaula@udg.es},
}
\author{J. Mart\'{\i}}{
  address={Departamento de F\'{\i}sica, Escuela Polit\'ecnica Superior, Univ. de Ja\'en, Virgen de la Cabeza 2, 23071 Ja\'en, Spain},
  email={jmarti@ujaen.es},
}

\copyrightyear  {2001}

\begin{abstract}
LS~I~+61~303 is one of the most studied X-ray binary systems because of its
two peculiarities: On the one hand being the probable counterpart of the 
variable gamma ray source 2CG~135+01 \cite{gregory:1978, tavani:1998} and on
the other hand being a periodic radio source \cite{taylor:1982}. The recent
discovery of a radio emitting jet extending ca. 200 AU at both sides of a
central core \cite{massiet:2004} in all evidence has shown the occurrence
of accretion/ejection processes in this system. However, the radio outbursts
do not occur at periastron passage, where the accretion is at its maximum, but
several days later. In addition, when the gamma-ray emission of 2CG~135+01 is
examined along the orbital phase of LS~I~+61~303 one sees that this emission
seems to peak at periastron passage \cite{massi:2004}. Here in detail
we analyse the trend of the gamma-ray data versus orbital phase and discuss the
delay between peaks at gamma-rays and in the radio band within the framework
of a two-peak accretion/ejection model proposed by Taylor et al.
\cite{taylor:1992} and further developed by Mart\'{\i} \& Paredes
\cite{marti:1995}.
\end{abstract}

\date{\today}

\maketitle

\section{The binary stellar system}

The binary stellar system LS~I~+61~303 has an orbital period of 26.4960 days
and a rather eccentric orbit ($e$=0.7) \cite{hutchings:1981, taylor:1982,
marti:1995, gregory:2002, casares:2004}. The visible companion is an early
type, rapidly rotating B0Ve star \cite{hutchings:1981, casares:2004}. Be stars
have a dense and slow ($<$100 km~s$^{-1}$) disk-like wind with a wind density
distribution following a power law; a long monitoring of the H$_{\alpha}$ line
emission has indicated that in the system LS~I~+61~303 are present variations
in the mass loss with a period of about 4.6 years \cite{zamanov:1999,
zamanov:2000, neish:2002}.

Through this dense, structured and variable envelope  the
compact object travels and accretes (see sketch in Fig.~1). It is still an open issue whether the
compact object in this system is a neutron star or a black hole. Recent
optical observations \cite{casares:2004} give a mass function $f$ in the range
$0.003<f<0.027$. The maximum value of $f$ for an inclination $i=38^{\circ}$
and a mass of the Be star $M = 18\,M_\odot$ implies a mass of the compact
object $M_{\rm X}<3.8\,M_\odot$; since the assumed upper limit for a stable
neutron star is $3\,M_\odot$, the presence of a black hole in the system
cannot be ruled out \cite{massi:2004}.

\begin{figure}
\includegraphics[height=.6\textheight, angle=-90]{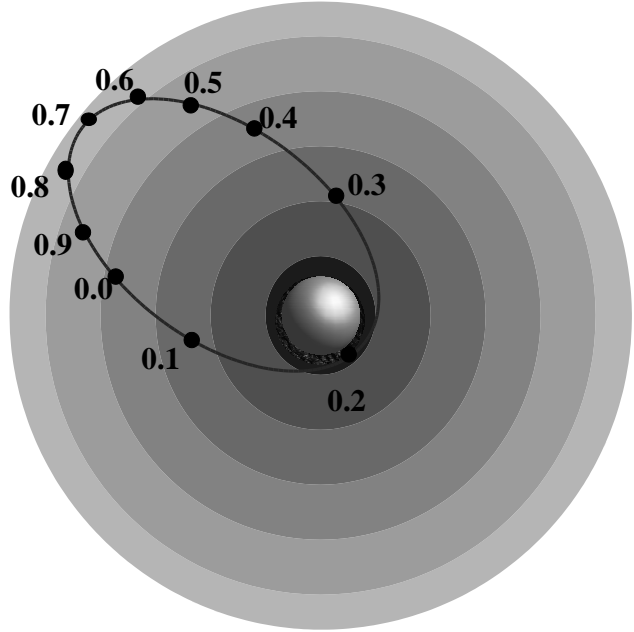}
\caption{Sketch of the binary system LS~I~+61~303: Orbital phases of the 
compact object travelling through the wind of the Be companion (see text).}
\label{sketch}
\end{figure}

\section {Radio emission}

One of the most unusual aspects of the radio emission is that it exhibits two
periodicities: a 26.4960 day periodic nonthermal outburst and a 4.6 years
modulation of the outburst peak flux \cite{taylor:1982, gregory:2002}. The
last period is clearly correlated with the mass loss of the Be star
\cite{zamanov:1999, zamanov:2000, neish:2002}, whereas the shorter periodicity
corresponds to the orbital period of the binary system \cite{hutchings:1981,
casares:2004}. The variations of the Be equatorial disk influence, besides the
outburst peak flux, the orbital phase where the outbursts occur and give rise
to the observed broad distribution $\phi_{\rm radio-outbursts}$=0.45--0.95
\cite{paredes:1990} where the phase is calculated for 
$t_0$~=~JD~2\,443\,366.775 and $P=26.4960$~days \cite{gregory:2002}.

As shown in Fig.~1 the $\phi$ at periastron passage is 0.2
\cite{casares:2004}. Therefore one of the fundamental questions concerning the
periodic radio outbursts of LS~I~+61~303 has been: If the maximum accretion
must occur where the density is highest, why are the radio outbursts delayed
with respect to the periastron passage?

On the other hand the radio emission has been resolved with VLBI, EVN and 
MERLIN \cite{massi:1993, massi:2001, massiet:2004, paredes:1998}. The 
morphology of the radio emission in Fig.~2a (at $\phi$=0.7) is  typical for
a microquasar, i.e. jet-like, and constitutes the observational evidence of
the occurrence of accretion/ejection processes in this system. Moreover, the
morphology in Fig.~2a is S-shaped and quite similar to that of the precessing
jet of SS~433 \cite{hjellming:1988}. The receding jet is attenuated by Doppler
de-boosting but still above the noise limit of the image. One day later
(Fig.~2b) the precession suggested in the first MERLIN image becomes evident:
a new feature is present oriented to the North-East at a position angle (PA)
of 67$^{\circ}$; the counter-jet is severely de-boosted. The
Northwest-Southeast jet of Fig.~2a has a PA=124$^{\circ}$. Therefore a quite
large rotation has occurred in only 24 hours.

\begin{figure}
\includegraphics[width=1.0\textwidth]{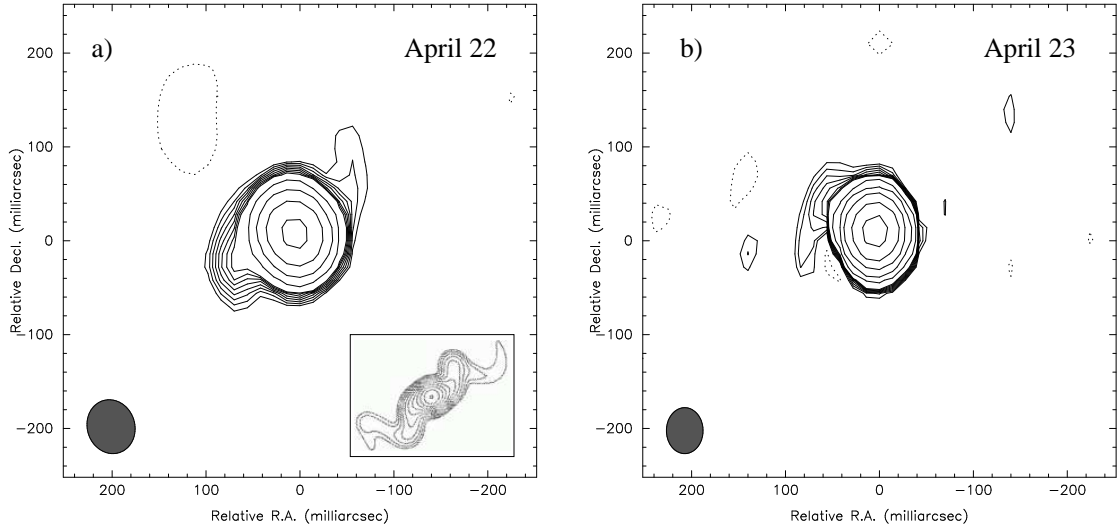}
\caption{
{\bf a)} MERLIN image of LS~I~+61~303 at 5~GHz obtained on 2001 April 22.
North is up and East is to the left. The synthesized beam has a size of
$51\times58$~mas, with a PA of 17$^{\circ}$. The contour levels are at $-$3,
3, 4, 5, 6, 7, 8, 9, 10, 20, 40, 80, and 160$\sigma$, being
$\sigma$=0.14~mJy~beam$^{-1}$. The S-shaped morphology strongly recalls the
precessing jet of SS~433, whose simulated radio emission (Fig.~6b in Hjellming
\& Johnston \cite{hjellming:1988}) is given in the small box. {\bf b)} Same as
before but for the April 23 run. The synthesized beam has a size of
$39\times49$~mas, with a PA of $-$10$^{\circ}$. The contour levels are the
same as those used in the April 22 image but up to 320$\sigma$, with
$\sigma$=0.12~mJy~beam$^{-1}$ \cite{massiet:2004}.}
\label{merlin}
\end{figure}

\section {Gamma-ray Emission}

In 1978 Gregory and Taylor \cite{gregory:1978} reported about the discovery of
a radio source (later on associated to LS~I~+61~303) within the 1$\sigma$
error circle of the COS~B $\gamma$-ray source 2CG~135+01. This association
remained however controversial because of the presence of the quasar
QSO~0241+622 within the relatively large COS B error box. The position of this
gamma-ray source given as 2EG~J0241+6119 in the Second EGRET Catalog is $l$ =
135$^{\circ}$.58, $b$ = 1$^{\circ}$.13; the radius of the 95\% confidence
error contour of about 13 arcminutes has finally ruled out the possible
identification with QSO~0241+622, 64 arcminutes away. The position is, on the
contrary compatible with LS~I~+61~303 only 8 arcminutes away
\cite{kniffen:1997}. In 1998 Tavani and collaborators establish the
possibility of variability of 2CG~135+01 on timescales of days
\cite{tavani:1998}. Massi \cite{massi:2004} examines the EGRET data as a function
of the orbital phase and notices the clustering of high flux around the $\phi$
interval 0.2--0.5 estimated to be the periastron passage.

Very recently Casares and collaborators \cite{casares:2004} have finally
confirmed periastron passage to be at $\phi$=0.2. Let us, therefore examine
the plot of Fig.~3. The plot begins at the established periastron passage
$\phi=$0.2 and shows the follow-up of the EGRET gamma-ray emission along one
full orbit. At epoch JD~2\,450\,334 (i.e. circles in the plot, with empty
circles indicating upper limits) the orbit has been well sampled at all
phases: A clear peak is centered at periastron passage 0.2 and 1.2. At a
previous epoch (JD~2\,449\,045; triangles in the plot) the sampling is
incomplete, still the data show an increase near periastron passage at $\phi
\simeq $0.3, and a peak at $\phi\simeq$0.5. The 3 squares refer to a third
epoch (JD~2\,449\,471).

\begin{figure}
\caption{EGRET data vs. orbital phase (see text).} 
\includegraphics[angle=-90, width=1.0\textwidth]{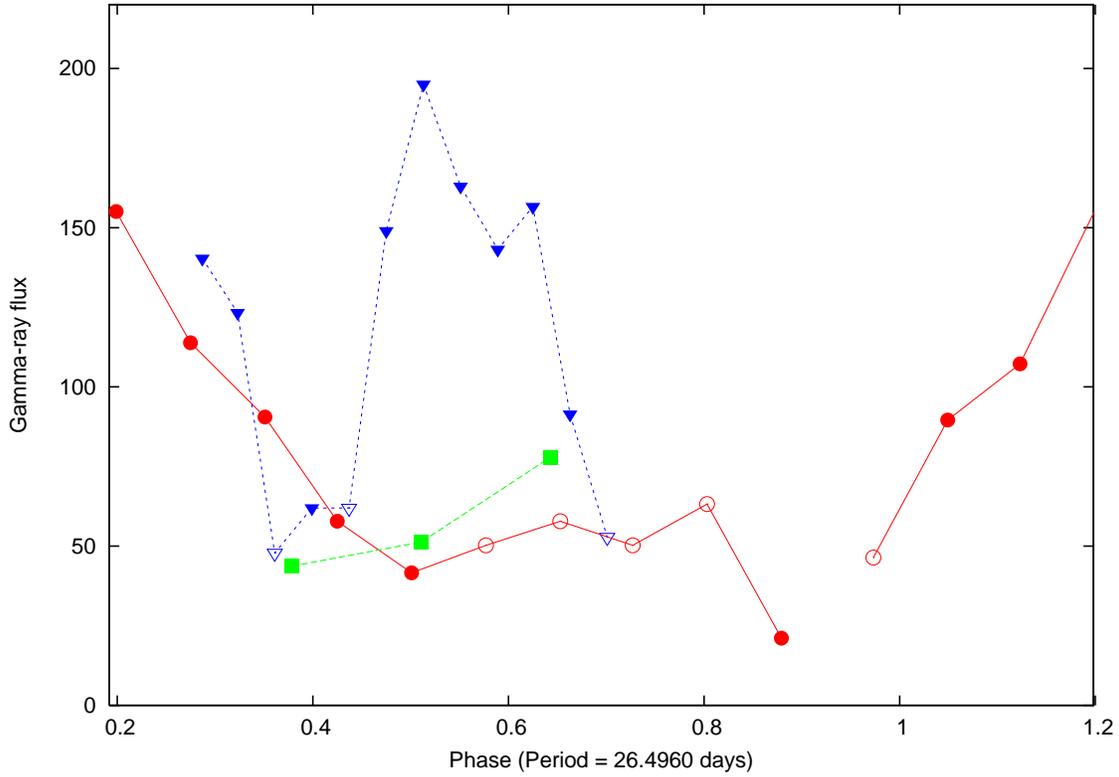}
\label{egret}
\end{figure}

\section {The two-peak accretion/ejection model}

Taylor et~al. \cite{taylor:1992} and Mart\'{\i} \& Paredes \cite{marti:1995}
have modelled the properties of this system in terms of an accretion rate
$\dot{M} \propto {\rho_{\rm wind}\over v_{\rm rel}^3}$, (where $\rho_{\rm
wind}$ is the density of the Be star wind and $v_{\rm rel}$ is the relative
speed between the accretor and the wind) which develops two peaks for high
eccentricities: the highest peak corresponds to the periastron passage
(highest density), while the second peak occurs when the drop in the relative
velocity $v_{\rm rel}$ compensates (because of the inverse cube dependence)
the decrease in density. Mart\'{\i} \& Paredes \cite{marti:1995} have shown
that both peaks are above the Eddington limit and therefore one expects that
matter is ejected twice within the 26.5 days interval. During the first
ejection, due to the proximity to the Be star, no radio emission but high
energy emission is expected because of severe inverse Compton losses.

As a matter of fact whereas the few gamma-ray data available show high levels
of gamma-ray emission around periastron passage (previous section), radio
outbursts were indeed never observed at periastron passage in more than 20
years of radio flux measurements. Moreover, Bosch-Ramon and Paredes
\cite{bosch:2004} have proposed a numerical model based on inverse Compton
scattering where the relativistic electrons in the jet are exposed to stellar
photons (external Compton) as well as to synchrotron photons (synchrotron self
Compton). This model considers accretion variations along the orbit and
predicts indeed a gamma-ray peak at periastron passage where the accretion is
higher.

At the second accretion peak the compact object is enough far away from the Be
star, so that the inverse Compton losses are small and electrons can propagate
out of the orbital plane. Then an expanding double radio source should be
observed, what in fact has been observed (Fig.~1) by MERLIN at orbital phase
0.7 (apoastron). Interesting in this respect is the gamma-ray peak at
$\phi\simeq0.5$ that could originate from a second ejection which occurred
still close enough to the Be star.

\section{Conclusions}

It is clear that new gamma-ray observations are desirable to confirm the 
periodical gamma-ray flares at periastron passage predicted by the model. For
a second ejection still close enough to the Be star in order to generate a
second gamma-ray flare, joint radio observations would allow to determine
important physical parameters. The radio and gamma-ray peaks are expected to
be delayed: When the electrons start to produce synchrotron emission in the
radio band, this emission is absorbed by the thermal electrons of the Be wind.
Such a delay strongly depends on the distance from the Be star.

In ejections distant enough from the Be star it could be verified if also disk
photons are upscattered by the relativistic electrons and  contribute to
the gamma-ray emission \cite{kaufman:2002}, such a contribution would finally
explain why LS~I~+61~303 is subluminous in the X-ray range (see Table~1,
adapted from \cite{combi:2004}).

\begin{table}
\begin{tabular}{ccc} 
\hline
  \tablehead{1}{c}{b}{$L_{\gamma}~(> \rm 100~MeV)$} &
  \tablehead{1}{c}{b}{$L_{\rm X}$} &
  \tablehead{1}{c}{b}{$L_{\rm radio}$} \\
\hline
3.1$\times$10$^{35}$     & (1--6)$\times$10$^{34}$ & (1--17)$\times$10$^{31}$\\
\hline
\end{tabular}
\caption{Luminosity (erg/s) \cite{combi:2004}}
\label{tab:b}
\end{table}

Moreover, on the basis of the precession shown in Fig.~2, we suggest that the
fast precession bringing the jet intermittently closer and farther to the
line of sight should produce noticeable variable $\gamma$-ray emission. In
fact, the amplification due to the Doppler factor for Compton scattering of
stellar photons by the relativistic electrons of the jet is even higher than
that for synchrotron emission \cite{kaufman:2002, massiet:2004}. In conclusion
LS~I~+61~303 becomes the ideal laboratory to test the recently proposed model
for microblazars with INTEGRAL and MERLIN observations now and by AGILE and
GLAST in the future.

\begin{theacknowledgments}
MERLIN is operated as a National Facility by the University of Manchester at Jodrell Bank Observatory on behalf of the UK Particle Physics \& Astronomy Research Council.
M.~R., J.~M.~P. and J.~M. acknowledge partial support by DGI of the Ministerio de Ciencia y Tecnolog\'{\i}a (Spain) under grant AYA2001-3092, as well as partial support by the European Regional Development Fund (ERDF/FEDER).
M.~R. acknowledges support by a Marie Curie Fellowship of the European Community programme Improving Human Potential under contract number HPMF-CT-2002-02053.
M.~P. acknowledges financial support by the program `Ram\'on y Cajal' of the
Ministerio de Ciencia y Tecnolog\'{\i}a (Spain).
J.~M. has been aided in this work by an Henri Chr\'etien International
Research Grant administered by the American Astronomical Society, and has been
partially supported by the Plan Andaluz de Investigaci\'on of the Junta de
Andaluc\'{\i}a (ref. FQM322).

\end{theacknowledgments}

\bibliographystyle{aipproc}   

\bibliography{1mmassiok}

\hyphenation{Post-Script Sprin-ger}
\begin{thebibliography}{8}
\expandafter\ifx\csname natexlab\endcsname\relax\def\natexlab#1{#1}\fi
\providecommand{\enquote}[1]{``#1''}
\expandafter\ifx\csname url\endcsname\relax
  \def\url#1{\texttt{#1}}\fi
\expandafter\ifx\csname urlprefix\endcsname\relax\def\urlprefix{URL }\fi

\bibitem[Brown and Austin(2000{\natexlab{b}})]{gregory:1978}
Gregory, P.C., \& Taylor, A.R.,
\emph{Nature}, \textbf{272}, 704--706 (1978).

\bibitem[Brown and Austin(2000{\natexlab{b}})]{tavani:1998}
Tavani, M., Kniffen, D., Mattox, J.R., Paredes, J.M., \& Foster, R.S.,
\emph{ApJ}, \textbf{497}, L89--L91 (1998).

\bibitem[Brown and Austin(2000{\natexlab{b}})]{taylor:1982}
Taylor, A.R., \& Gregory, P.C.,
\emph{ApJ}, \textbf{255}, 210--216 (1982).

\bibitem[Brown and Austin(2000{\natexlab{b}})]{massiet:2004}
Massi, M., Rib\'o, M., Paredes, J.M., Garrington, S.T., Peracaula, M., \& Mart\'{\i}, J.,
\emph{A\&A}, \textbf{414}, L1--L4 (2004).

\bibitem[Brown and Austin(2000{\natexlab{b}})]{massi:2004}
Massi, M.,
\emph{A\&A}, \textbf{422}, 267--270 (2004).

\bibitem[Brown and Austin(2000{\natexlab{b}})]{taylor:1992}
Taylor, A.R., Kenny, H.T., Spencer, R.E., \& Tzioumis, A.,
\emph{ApJ}, \textbf{395}, 268--274 (1992).

\bibitem[Brown and Austin(2000{\natexlab{b}})]{marti:1995}
Mart\'{\i}, J., \& Paredes, J.M.,
\emph{A\&A}, \textbf{298}, 151--158 (1995).

\bibitem[Brown and Austin(2000{\natexlab{b}})]{hutchings:1981}
Hutchings, J.B., \& Crampton, D.,
\emph{PASP}, \textbf{93}, 486--489 (1981).

\bibitem[Brown and Austin(2000{\natexlab{b}})]{gregory:2002}
Gregory, P.C.,
\emph{ApJ}, \textbf{575}, 427--434 (2002).

\bibitem[Brown and Austin(2000{\natexlab{b}})]{casares:2004}
Casares, J., Ribas, I., Paredes, J.M., Mart\'{\i}, J., \& Allende Prieto, C.,
\emph{MNRAS}, submitted (2004).

\bibitem[Brown and Austin(2000{\natexlab{b}})]{zamanov:1999}
Zamanov, R.K., Mart\'{\i}, J., Paredes, J.M., Fabregat, J., Rib\'o, M., \& Tarasov, A.E.,
\emph{A\&A}, \textbf{351}, 543--550 (1999).

\bibitem[Brown and Austin(2000{\natexlab{b}})]{zamanov:2000}
Zamanov, R.K., \& Mart\'{\i}, J.,
\emph{A\&A}, \textbf{358}, L55--L58 (2000).

\bibitem[Brown and Austin(2000{\natexlab{b}})]{neish:2002}
Gregory, P.C., \& Neish, C.,
\emph{ApJ}, \textbf{580}, 1133--1148 (2002).

\bibitem[Brown and Austin(2000{\natexlab{b}})]{paredes:1990}
Paredes, J.M., Estalella, R. \& Rius, A.,
\emph{A\&A}, \textbf{232}, 377--380 (1990).

\bibitem[Brown and Austin(2000{\natexlab{b}})]{massi:1993}
Massi, M., Paredes, J.M., Estalella, R., \& Felli, M.,
\emph{A\&A}, \textbf{269}, 249--254 (1993).

\bibitem[Brown and Austin(2000{\natexlab{b}})]{massi:2001}
Massi, M., Rib\'o, M., Paredes, J.M., Peracaula, M., \& Estalella, R.,
\emph{A\&A}, \textbf{376}, 217--223 (2001).

\bibitem[Brown and Austin(2000{\natexlab{b}})]{paredes:1998}
Paredes, J.M., Massi, M., Estalella, R., \& Peracaula, M.,
\emph{A\&A}, \textbf{335}, 539--544 (1998).

\bibitem[Brown and Austin(2000{\natexlab{b}})]{hjellming:1988}
Hjellming, R.M., \& Johnston, K.J.,
\emph{ApJ}, \textbf{328}, 600--609 (1988).

\bibitem[Brown and Austin(2000{\natexlab{b}})]{kniffen:1997}
Kniffen, D.A., Alberts, W.C.K., Bertsch, D.L., Dingus, B.L., Esposito, J.A., et~al.,
\emph{ApJ}, \textbf{486}, 126--131 (1997).

\bibitem[Brown and Austin(2000{\natexlab{b}})]{bosch:2004}
Bosch-Ramon, V., \& Paredes, J.M.,
\emph{A\&A}, \textbf{425}, 1069--1074 (2004).

\bibitem[Brown and Austin(2000{\natexlab{b}})]{kaufman:2002}
Kaufman Bernad\'o, M.M., Romero, G.E., \& Mirabel, I.F.,
\emph{A\&A}, \textbf{385}, L10--L13 (2002).

\bibitem[Brown and Austin(2000{\natexlab{b}})]{combi:2004}
Combi, J.A., Rib\'o, M., Mirabel, I.F., \& Sugizaki, M.,
\emph{A\&A}, \textbf{422}, 1031--1037 (2004).

\end{thebibliography}

\IfFileExists{\jobname.bbl}{}
 {\typeout{}
  \typeout{******************************************}
  \typeout{** Please run "bibtex \jobname" to optain}
  \typeout{** the bibliography and then re-run LaTeX}
  \typeout{** twice to fix the references!}
  \typeout{******************************************}
  \typeout{}
 }

\end{document}